\documentclass[fleqn,twoside]{article}
\usepackage{espcrc2}
\usepackage{graphicx}
\usepackage{amssymb}

\newcommand{\be}{\begin{eqnarray}}
\newcommand{\ee}{\end{eqnarray}}
\newcommand{\el}{\nonumber\\}
\newcommand{\1}{\mathbb{I}}
\newcommand{\tr}{\mathrm{Tr}}
\newcommand{\dt}{\mathrm{Det}}
\newcommand{\p}{\partial}
\newcommand{\ex}[1]{\mathrm{e}^{#1}}
\newcommand{\conj}[1]{{{#1}^{*}}}
\newcommand{\dslash}{\makebox[0cm][l]{$\,/$}D}

\title{Eigenvalue correlations in QCD with a chemical potential}

\author{James C. Osborn\address{Physics Department, Boston University,
 Boston, MA 02215, USA}}

\begin{document}

\begin{abstract}
We discuss a new Random Matrix Model for QCD with a chemical potential
that is based on the symmetries of the Dirac operator and can be
solved exactly for all eigenvalue correlations for any number of
flavors.  In the microscopic limit of small energy levels the results
should be an accurate description of QCD.  This new model can also be
scaled so that all physical observables remain at their $\mu=0$ values
until a first order chiral restoration transition is reached.  This
gives a more realistic model for the QCD phase diagram than previous
RMM.  We also mention how the model might aid in determining the phase
diagram of QCD from future numerical simulations.
\end{abstract}

\maketitle

It is well known that at zero chemical potential the low energy modes
of QCD can be described by a chiral effective Lagrangian
\be
{\cal L} = 
   \frac{F^2}{4} \,\tr \left[\,\p_\mu U \,\p_\mu U^\dagger \right]
 - \frac{\Sigma}{2} \,\tr \left[\, M \,(U + U^\dagger ) \right] ~.
\ee
In a finite volume and at low energy when
$1/{m_\pi} \gg L$ (also known as the $\epsilon$-regime)
the kinetic term can be ignored and the theory reduces to simply
an integration over the zero momentum part of the SU($N_f$)
matrix $U$ that parameterizes the Goldstone manifold.
The results then do not depend on the dynamical details of QCD but
rely solely on the symmetries.
In this limit the chiral effective theory can also be expressed as a
Random Matrix Theory (RMT) with the appropriate
chiral symmetry \cite{chrmt}.
The partition function is given in terms of an integral over all
$(N+\nu) \times N$ complex matrices $A$
\be
Z = \int dA ~ \ex{- \sigma \, N \, \tr \; A^\dagger A} ~
 \dt({\cal D}+m)^{N_f}
\ee
with a Dirac matrix
\be
{\cal D} + m =
\left(
 \begin{array}{cc}
  m~ \1    & i A \\
  i A^\dagger & m~ \1 \\
 \end{array}
\right) ~.
\ee
The choice of a complex matrix corresponds to a SU($N_c\ge3$)
gauge field with $\nu$ the topological charge.
In the microscopic limit of $N$ going to infinity with $N m$ held
fixed the chiral RMT is equivalent to the static part of the
chiral effective Lagrangian.
The RMT form of the effective theory, however, is better suited for
studying the eigenvalues of the Dirac operator.

Here we want to examine the extension of the chiral RMT to include
a baryon chemical potential.
In this case a Hermitian term $\mu \gamma_0$ is added to the
anti-Hermitian $\dslash$.
The eigenvalues of the Dirac operator become spread out in the
complex plane and this gives
rise to a sign problem that hinders numerical investigations.

One model considered previously by Stephanov \cite{S}
used the Dirac matrix
\be
\mathcal{D} = \left( \begin{array}{cc}
 0  & i A+\mu \,\1 \\
 i A^\dagger + \mu \,\1  & 0 
 \end{array} \right)
\ee
where the chemical potential $\mu$ is multiplied by an identity matrix.
This has the correct symmetries but the fixed form of the chemical potential term
makes it very difficult to calculate the eigenvalues analytically.
Also it has been shown that this model has an unphysical $\mu$ dependence
when it is studied at larger $\mu$ \cite{unp}.

\begin{figure}[t]
\includegraphics[width=\columnwidth]{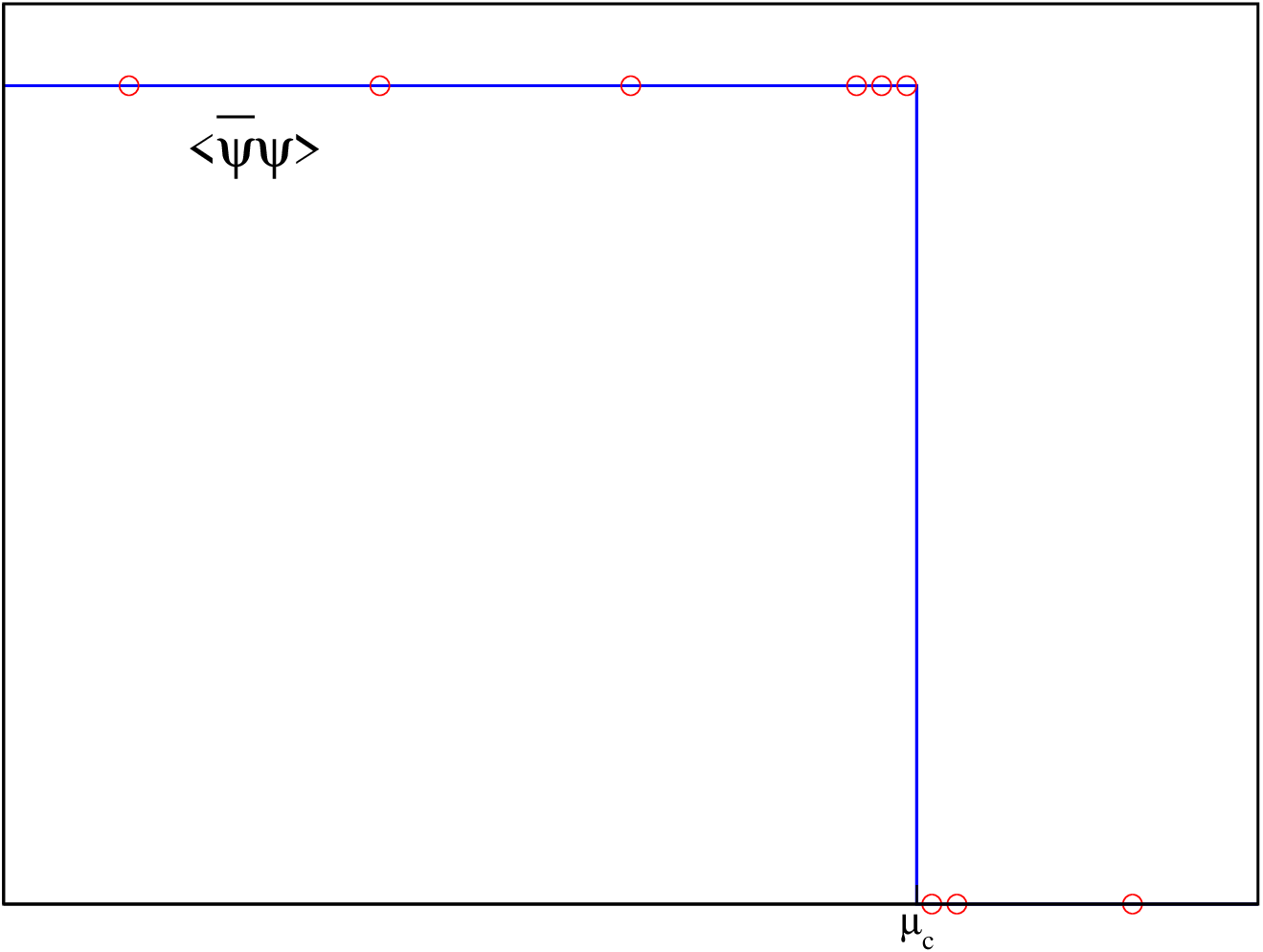}
\vspace{-14mm}
\caption{Example of a direct measurement of a first order chiral phase transition.
To determine $\mu_c$ accurately one needs several data points around $\mu_c$.}
\vspace{-5mm}
\label{figcc}
\end{figure}

A new model for QCD with a chemical potential was introduced recently \cite{O}.
This uses a different form for the Dirac matrix
\be
\mathcal{D} = \left( \begin{array}{cc}
0 & i A + \mu  B \\
i A^{\dagger} + \mu B^{\dagger} & 0
\end{array} \right)
\label{dnew}
\ee
where the chemical potential part is now also modeled with a Gaussian
random matrix $B$.
This model still has the correct symmetries and can be solved
for all eigenvalue correlations analytically.
Additionally we can remove the unphysical $\mu$ dependence at zero temperature
giving a more physical model for chiral symmetry breaking with a chemical
potential.

\begin{figure}[t]
\includegraphics[width=\columnwidth]{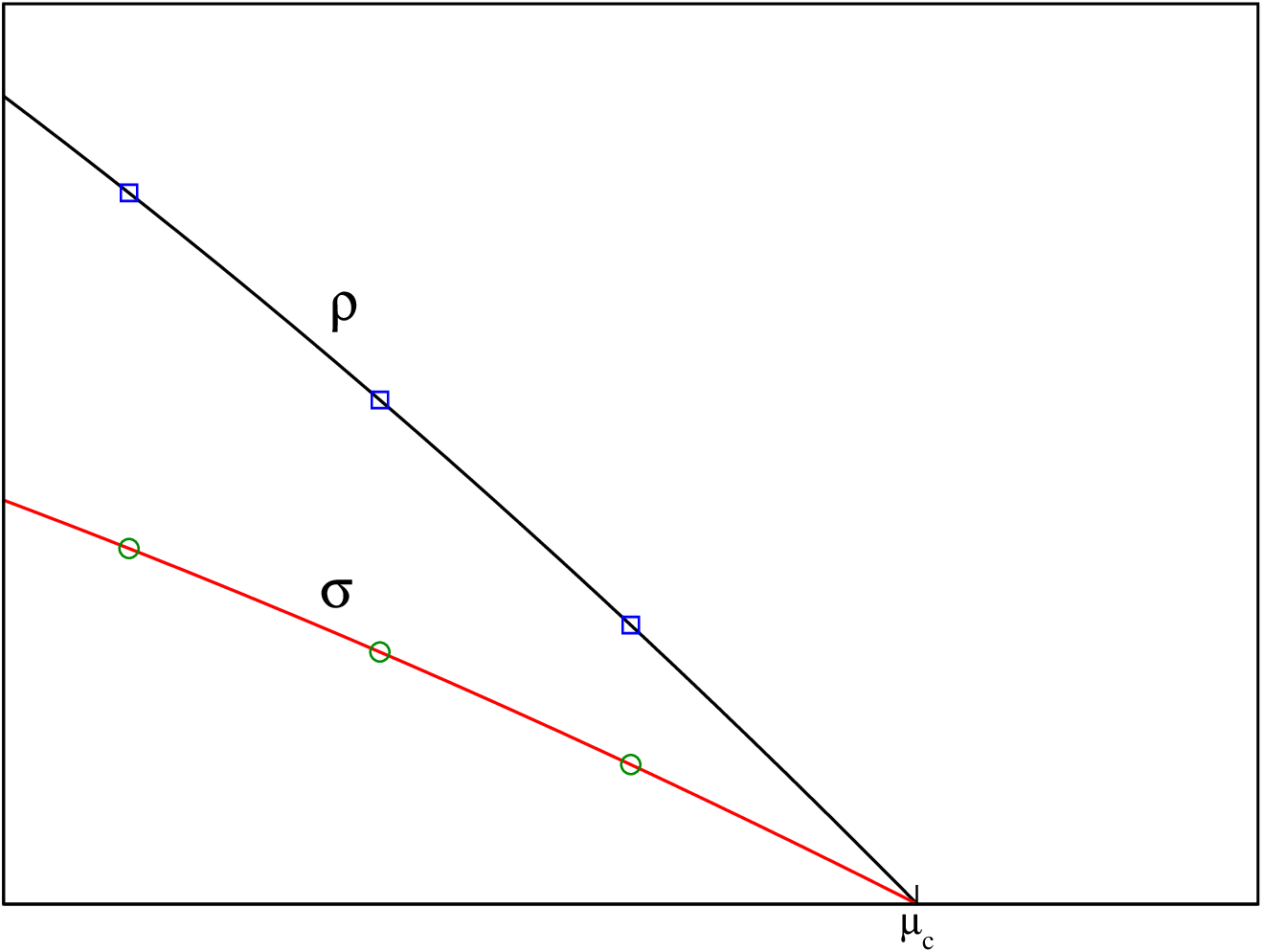}
\vspace{-14mm}
\caption{By extrapolating the effective parameters $\sigma$ and $\rho$ to zero 
one could determine $\mu_c$ at a first order phase transition without
direct simulations at larger chemical potentials.}
\vspace{-5mm}
\label{figab}
\end{figure}

By choosing an appropriate parameterization for the matrices $A$ and $B$ and
then integrating out all angular variables the
new model can be written in terms of the complex eigenvalues $z_k$ 
of (\ref{dnew}) as \cite{O}
\be
\label{epfnew}
Z \propto \prod_{j=1}^{N_f} m_j^{\nu}
 \int \left|\Delta(z^2)\right|^2
 \prod_{k=1}^{N} w(z_k) dz_k
 \prod_{j=1}^{N_f} (z_k^2 + m_j^2)
\ee
with the Vandermonde determinant
\be
\Delta(z^2) = \prod_{i<j} (z_i^2-z_j^2)
\ee
and a weight function containing a modified Bessel function
\be
\label{wnew}
w(z) &=& |z|^{2\nu+2}
\exp\left( \frac{\sigma N (1-\mu^2)}{4 \mu^2} (z^2 + \conj{z}^2) \right) \el
&\times& K_\nu \left( \frac{\sigma N (1+\mu^2)}{2 \mu^2} |z|^2 \right) ~.
\ee
This representation makes it easy to solve for all eigenvalue correlations
for any number of flavors using the orthogonal polynomial method.
The results are given in \cite{O} which we won't reproduce here.
In the microscopic limit the eigenvalue correlations from the RMT are
expected to agree with QCD due to universality.
Recent lattice studies support universality in the eigenvalue
density at small chemical potentials \cite{univlat}.

We now examine the behavior of this model at larger chemical potentials.
If we look, for example, at the result for the partition function for
one flavor given by a Laguerre polynomial
\be
%\hspace{-1cm}
Z_{N_f=1} \propto
 m^\nu L_{N}^\nu \left(\frac{- \sigma N m^2}{ 1-\mu^2}\right)
\ee
we see that the $\mu$ dependence comes in only through the combination
$\sigma/(1-\mu^2)$.
At $T=0$ we expect no $\mu$ dependence of thermodynamic quantities
until a first order phase transition is reached.
We can remove the $\mu$ dependence completely by setting
$\sigma(\mu) = \sigma_0 \, (1 - \mu^2)$.
We then find that all thermodynamic observables such as the
chiral condensate in the chiral limit
\be
\langle \bar\psi \psi \rangle = 
 \frac{2N}{V} \sqrt{\frac{\sigma}{1-\mu^2}} = 
 \frac{2N}{V} \sqrt{\sigma_0}
\ee
now do not depend on the chemical potential until
a phase transition is reached.
This therefore provides a reasonable effective model
for QCD with a chemical potential at zero temperature.

We can also use this as an effective model at nonzero temperature.
To do this we first include an explicit scale for $\mu$ in the
Gaussian weight for the matrix $B$ so that the matrix weights are
\be
\exp(- \, \sigma \, N \, \tr \;[ A^\dagger A ] 
     - \, \rho \, N \, \tr \;[ B^\dagger B ] \,) ~.
\ee
This gives the same results as replacing $\mu$ with
$\mu\sqrt{\sigma/\rho}$ in previous expressions.
The chiral condensate in the chiral limit is then
\be
\langle \bar\psi \psi \rangle = 
 \frac{2N}{V} \sqrt{\frac{\sigma\; \rho}{\rho-\sigma\, \mu^2}} ~.
\ee
From this we see that there are two possible scenarios for the
chiral condensate to vanish as $\mu$ is increased.
One is that $\sigma$ goes to zero while $\rho$ remains nonzero.
This will be first order only if $\sigma$ has a discontinuous
jump to zero, otherwise it is second order.
Another possibility is that both $\sigma$ and $\rho$ reach zero at the same
point.  In this case the transition can be first order even if
$\sigma$ and $\rho$ are continuous.

The results of the effective model may be very useful to future
numerical studies.
For example in order to locate the possible first order chiral phase
transition at large $\mu$ with a direct simulation one would need
several measurements of the condensate near the transition
as illustrated in figure \ref{figcc}.
This is after performing the necessary thermodynamic extrapolation
to find a sharp transition.
Due to the severe sign problem, performing the necessary simulations
at such a large $\mu$ is likely to remain prohibitive for a long time.
An alternative approach is to measure the effective parameters
$\sigma$ and $\rho$ at smaller $\mu$ and then attempt to extrapolate
them to zero.
The critical point is given by when $\sigma$ reaches zero.
If $\rho$ also reaches zero at the same spot then it could be a
first order transition.
This allows one to find the critical point without going to
very large $\mu$.
Of course current simulations still have difficulty at the moderate
$\mu$ needed to extrapolate, so it will likely still be a while before
the numerical simulations are feasible.

\begin{figure}[t]
\includegraphics[width=\columnwidth,height=\columnwidth]{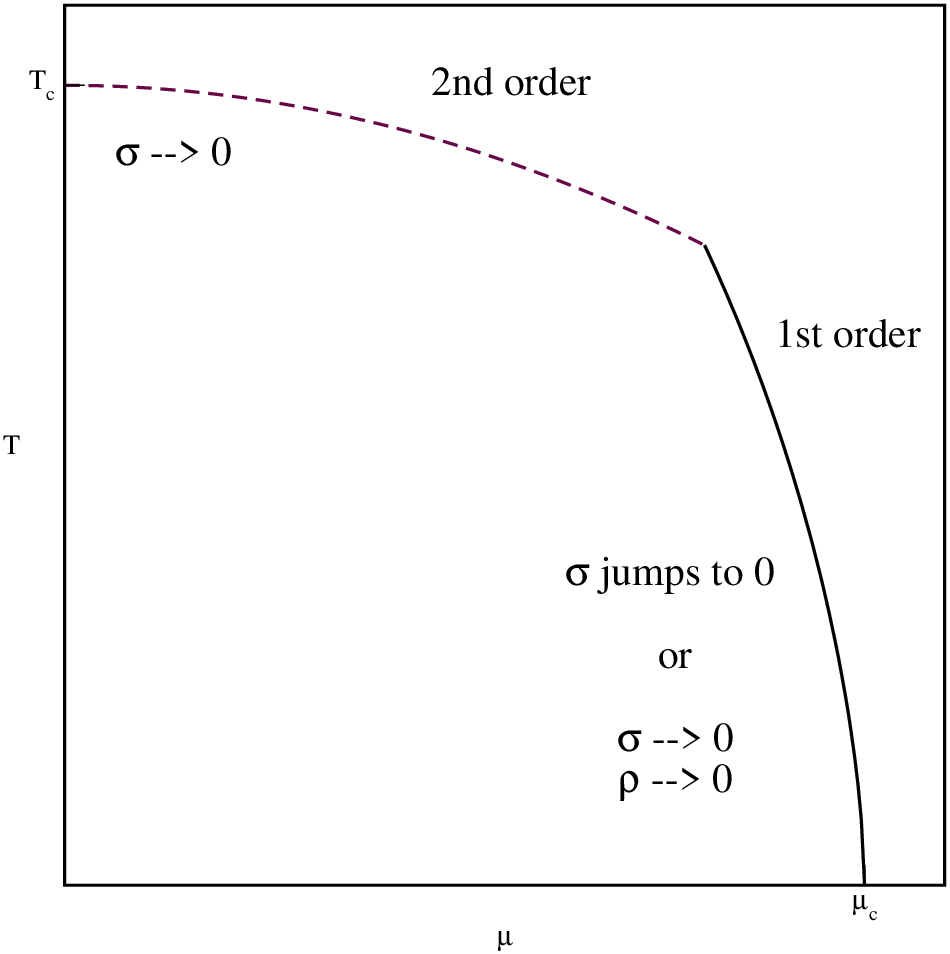}
\vspace{-14mm}
\caption{Possible phase diagram.}
\vspace{-5mm}
\label{figphase}
\end{figure}

The most general form of effective theory for chiral symmetry breaking
with a baryon chemical potential can be obtained by
letting $\sigma(T,\mu)$ and $\rho(T,\mu)$ be functions of $T$ and $\mu$.
The parameters $\sigma$ and $\rho$ can be determined from numerical
simulations by fitting to a quantity such as the eigenvalue density
or the valence quark mass dependence of the chiral condensate.
The latter is likely to be a better observable since it is smoother
and should be easier to obtain with higher statistics.
Then by mapping out the parameters in the $T$-$\mu$ plane one
could end up with a phase diagram like the example in figure
\ref{figphase}.

We now have a new Random Matrix Model for QCD with a chemical potential
that can be solved for all eigenvalue correlations for any number
of flavors.
In the microscopic limit this provides exact expressions that
should agree with finite density QCD due to expected universality.
This model can also be used as an effective theory for
chiral symmetry breaking in the presence of a chemical potential.
By considering the effective parameters in the model to be general
functions of $T$ and $\mu$ this may provide a useful way
to help map out QCD phase diagram.

\end{document}